\documentclass[twocolumn,showpacs,preprintnumbers,amsmath,amssymb,prl,superscriptaddress]{revtex4}

\usepackage{amsmath,amssymb,amsthm,color}
\usepackage{graphicx}
\usepackage{dcolumn}

\newcommand {\beq}{\begin{eqnarray}}
\newcommand {\eeq}{\end{eqnarray}}

\begin{document}

\preprint{CALT  68-2897}

\title{Does anomalous violation of null energy condition invalidate holographic $c$-theorem?}

\author{Yu Nakayama}

\affiliation{California Institute of Technology, 452-48, Pasadena, California 91125, USA}


\begin{abstract}
Null energy condition plays a crucial role in holographic renormalization group flow, leading to the holographic $c$-theorem. Unfortunately, the null energy condition is quantum mechanically violated. Even the averaged version can be violated.
We discuss how the anomalous violation of the null energy condition affects the holographic renormalization group flow in $1+3$ dimensional bulk gravity. We show that despite the violation of the null energy condition, a suitably modified holographic $c$-function with a peculiar log correction is still monotonically decreasing in so far as we add the counterterm that removes a ghost mode of gravity.
\end{abstract}

\maketitle

\section{1. Introduction}
The holographic renormalization is a beautiful scheme to geometrize the renormalization group flow. In the classical Einstein gravity approximation, the Weyl consistency condition of the local renormalization group is automatically realized  as gravitational equations of motion. In particular, we can derive the holographic $c$-theorem \cite{Girardello:1998pd}\cite{Freedman:1999gp}\cite{Myers:2010xs}\cite{Myers:2010tj} that dictates there exists a $c$-function that monotonically decreases along the renormalization group flow in any space-time dimension.

In order to derive the holographic $c$-theorem, one crucial assumption is that the matter satisfies the null energy condition. This is reasonable at least classically because the field theoretic $c$-theorem, whose complete proof is available in $(1+1)$ dimension \cite{Zamolodchikov:1986gt} with recent significant progress in $(1+3)$ dimension \cite{Komargodski:2011vj}, requires the unitarity as a part of the assumption, and the null energy condition is naturally regarded as its gravitational counterpart. 

In general relativity, any pathological space-time could be a solution of the equation of motion without assuming the energy condition because we can simply declare that the corresponding Einstein tensor is sourced by the same energy-momentum tensor. It is known that the null energy condition, reasonably satisfied in realistic classical matter systems, is sufficiently strong to avoid many ``pathological" space-times such as wormhole, superluminal propagation, time-machine, shrinking black holes and so on.

Unfortunately (or fortunately), the null energy condition is violated quantum mechanically. Actually, the violation of the null energy condition is rather crucial for the consistency of quantum gravity in various ways. The Hawking radiation violates the classical area non-decreasing theorem proved by the null energy condition.  The violation of the null energy condition in $(1+1)$ dimensional worldsheet makes graviton massless in string theory. Orientifolds as they are also break the null energy condition, but they are crucial ingredients in string dualities (see e.g. \cite{Marolf:2002np} for a study how objects with the negative tension will affect the second law in string theory).

To relax the null energy condition while still avoiding pathological space-time, there have been various modifications proposed. One promising  direction is to average over the null geodesics. The so-called averaged null energy condition seems to hold in any quantum states in Minkowski space-time \cite{Klinkhammer:1991ki}\cite{Wald:1991xn}\cite{Graham:2007va},
but the violations were reported in curved backgrounds  \cite{Visser:1994jb}\cite{Urban:2009yt}\cite{Urban:2010vr} (technically we have to restrict ourselves to the achronal average because in compact space-time the averaged null energy condition is easily violated due to Casimir energy: in any way, even the achronal version is violated). The violation is either induced by quantum states or anomalous contributions in the energy-momentum tensor induced by trace anomaly in curved space-time.

Do reported violations of the null energy condition invalidate the holographic $c$-theorem? Unlike the favored violation we mentioned, we do not want the violation to kill the holographic $c$-theorem because we believe that $c$-theorem is universally true in dual unitary relativistic quantum field theories. Most of the reported violations are not immediately threatening to us because the holographic renormalization group flow takes place in a boundary Poincar\'e invariant setup, and we do not consider time-dependent non-vacuum process. However, there are known possibilities of violating null energy condition in a vacuum setup from an anomalous contribution to the energy-momentum tensor in curved space-time due to the trace anomaly. It is a universal violation in any theory and it has been used to violate even the averaged  null energy condition \cite{Visser:1994jb}\cite{Urban:2009yt}\cite{Urban:2010vr}.

In this paper, we would like to address the question and answer in a positive way. With a further thought, the violation of the (averaged) null energy condition are not important unless they appear in the self-consistent background that solves the quantum modified gravity equation. Indeed, there has been no reported violation of the averaged null energy condition in the self-consistent background \cite{Penrose:1993ud}\cite{Flanagan:1996gw}\cite{Graham:2007va}. We will see that a suitably modified holographic $c$-function with a peculiar log correction is still monotonically decreasing in so far as we add the counterterm that removes a ghost mode of gravity. This can be regarded as a consequence of the self-consistency.

The organization of the paper is as follows. In section 2, we review the holographic renormalization group and holographic $c$-theorem in $(1+3)$ dimension with possible higher derivative corrections. In section 3, we discuss the anomalous contribution to the energy-momentum tensor and possible violation of the null energy condition to see the fate of the holographic $c$-theorem. In section 4, we summarize our findings and discuss possible future directions.

\section{2. Holographic Renormalization Group and $c$-theorem}
As a starting point of our discussion, let us consider the holographic renormalization group flow and the holographic $c$-theorem in $(1+3)$ dimensional Einstein gravity\begin{align}
 S = \frac{1}{2} \int d^4x \sqrt{-g} \left(R + L_{\mathrm{matter}} \right) \ .
\end{align}
Throughout the paper, the Planck length is set to be one. In holographic renormalization group flow, we will consider the asymptotically AdS space-time whose metric is
\begin{align}
ds^2 = dr^2 + e^{2A(r)} \eta_{\mu\nu} dx^\mu dx^\nu \ , \label{holoren}
\end{align}
where $\eta_{\mu\nu} = (-1,+1,+1)$ is the three-dimensional flat Minkowski space-time metric, and $A(r) \to A_{\mathrm{UV}} r $ as $r \to +\infty$ and $A(r) \to A_{\mathrm{IR}} r$ as $r \to -\infty$ for the flow between two dual conformal field theories.

The holographic $c$-function, denoted by $a(r)$ for a conventional reason, is defined by
\begin{align}
a(r) \equiv  \frac{\pi^{3/2}}{\Gamma(3/2) (A'(r))^2} \ , 
\end{align}
where $A'(r) = \frac{dA(r)}{dr}$. At the fixed point $r \to \pm \infty$ it was interpreted as the universal term in the entanglement entropy of the dual conformal field theory.
By using the Einstein equation, one can compute the change of the holographic $c$-function along the holographic renormalization group flow as
\begin{align}
a'(r) &= -\frac{2\pi^{3/2}}{\Gamma(3/2) (A'(r))^3} A''(r) \cr
      &= - \frac{\pi^{3/2}}{\Gamma(3/2) (A'(r))^3} (T^{t}_{\ t} - T^{r}_{\ r}) \ge 0 \ , \label{deriv}
\end{align}
where $T_{\mu\nu}$ is the matter energy-momentum tensor, and the last inequality is the claimed holographic $c$-theorem. To justify the inequality, we have assumed that the null energy condition is satisfied so that $T^{r}_{\ r} - T^{t}_{\ t} \ge 0$.

The null energy condition demands that for any null vector $k_\mu$ such that $k_\mu k^\mu = 0$, the energy-momentum tensor must satisfy the inequality $T_{\mu\nu} k^\mu k^\nu \ge 0$. In our example, we choose $k^\mu = (1,e^{-A(r)},0,0)$. We will discuss the ``normalization" of the null vector later when we discuss the averaged condition, but it is irrelevant here. The null energy condition leads to $-A''(r) \ge 0$ in the holographic renormalization group flow, and it has played a crucial role in establishing the holographic $c$-theorem.

In addition to the null energy condition, there was an implicit technical assumption $A'(r) \ge 0$ in \eqref{deriv}. This can be derived from the fact that $A'(r \to \pm \infty) >0$ and $A''(r) \le 0$ from the null energy condition \cite{Myers:2010tj}.

Before introducing higher derivative corrections, we have a couple of comments here. The first observation, which will be useful later, is that the metric for the holographic renormalization \eqref{holoren} is conformally flat and the Weyl tensor $W_{\mu\nu\rho\sigma}$ vanishes. This is tightly related to the fact that our holographic renormalization group flow preserves the boundary Poincar\'e invariance.

The second comment is that given the recent success in proving the weak version of the $c$-theorem in $(1+3)$ dimensioanl quantum field theories \cite{Komargodski:2011vj} (with the lack of a non-perturbative proof of the strong version \cite{Cardy:1988cwa}\cite{Jack:1990eb}), one may be tempted to only require the weak version of the holographic $c$-theorem (note, however, we are dealing with the  $(1+2)$ dimensional boundary, where things are less clear). 
This is closely related to the averaged null energy condition. The averaged null energy condition only demands that $\int_\gamma T_{\mu\nu} k^\mu k^\nu d\lambda \ge 0$ over any (achronal) null geodesics $\gamma$ with the affine parameter $\lambda$ (such that $\partial_\mu \lambda k^\mu = 1$). As mentioned in the introduction, the averaged null energy condition is more difficult to violate than point-wise null energy condition. In any way, in our holographic $c$-thereom, we needed a slightly different averaging: $\int_\gamma T_{\mu\nu} k^\mu k^\nu f(\lambda) d\lambda \ge 0$, where $\lambda = r$ and $f(\lambda) = \frac{1}{(A'(r))^3}$ in order to show $a_{\mathrm{UV}} - a_{\mathrm{IR}} \ge 0$. Although we will focus on the strong $c$-theorem in the following, it would be interesting to understand the physical origin of this averaging.

The final comment is on the relation between holographic $c$-theorem and the holographic equivalence of scale invariance and conformal invariance. For the holographic equivalence to work, we need to assume a strict version of the null energy condition that demands that the matter must be trivial when the null energy condition is identically saturated \cite{Nakayama:2010wx}\cite{Nakayama:2010zz}. With this respect, we recall that the null energy condition is not enough to exclude the pathological situation where the matter has zero kinetic energy, and in order to guarantee the unitarity, it is not sufficient. Whenever the null energy condition is violated, the statement of the strict null energy condition is obscure. Fortunately, due to the symmetry of the problem, the anomalous violation we will discuss in the next section does not play an important role there.

Now let us introduce higher derivative corrections to the holographic $c$-theorem argument \cite{Myers:2010tj}. We generalize the Einstein-Hilbert action with various curvature squared terms:
\begin{align}
S =& \frac{1}{2} \int d^4x \sqrt{g}\left(\frac{6}{L^2} \alpha + R + \right. \cr
&+  L^2 \left(\lambda_1 R_{\mu\nu\rho\sigma} R^{\mu\nu\rho\sigma} + \lambda_2 R_{\mu\nu}R^{\mu\nu} + \lambda_3 R^2 \right. \cr
&+\left.\left. \tilde{\lambda} \epsilon^{\alpha\beta \gamma \delta} R_{\alpha \beta \rho \sigma} R_{\gamma \delta}^{\ \ \rho \sigma} \right) \right) \ . \label{higher}
\end{align}
Although for completeness we have added the parity odd Hirzebruch-Pontryagin term $\tilde{\lambda} \epsilon^{\alpha\beta \gamma \delta} R_{\alpha \beta \rho \sigma} R_{\gamma \delta}^{\ \ \rho \sigma}$, it is topological, and does not affect the bulk holographic renormalization group flow. Similarly, a particular combination of the parity even term with $(\lambda_1,\lambda_2,\lambda_3) = (1,-4,1)\lambda$ gives the Euler density: $(\mathrm{Euler} = R_{\mu\nu\rho\sigma}^2 -4R_{\mu\nu}^2+ R^2 $), so it  will not affect the bulk holographic renormalization group flow, either. Therefore, without losing generality, we can set $\lambda_1 = \tilde{\lambda} = 0$. 

The modification of the equations of motion is obtained from the metric variation
\begin{align}
I_{\mu\nu} &= (-g)^{-1/2} \frac{\delta}{\delta g^{\mu\nu}} \int \sqrt{-g} d^4 x R^2 \cr
&= 2D_{\mu} D_{\nu} R - 2 RR_{\mu\nu} + \left(\frac{1}{2}R^2 -2 \Box R\right) g_{\mu\nu}\cr
J_{\mu\nu} &= (-g)^{-1/2} \frac{\delta}{\delta g^{\mu\nu}} \int \sqrt{-g} d^4 x R_{\alpha\beta} R^{\alpha\beta} \cr
&= D_{\mu}D_\nu R - \Box R_{\mu\nu} - 2R^{\beta}_{\alpha} R^{\alpha}_{\ \mu \beta \nu} \cr 
&+ \left(\frac{1}{2}R_{\alpha\beta} R^{\alpha\beta} - \frac{1}{2} \Box R \right) g_{\mu\nu} \ .
\end{align}
We note that $I_{\mu\nu} = 3J_{\mu\nu}$ for conformally flat metric (actually, this is true more generally for any conformally Einstein metric). Thus, in the holographic renormalization group flow, there is only one parameter out of four within the curvature squared corrections that will eventually affect the flow of the holographic $c$-function.

The strategy to obtain a monotonically decreasing holographic $c$-function in higher derivative gravity advocated in \cite{Myers:2010tj} is to demand that only $A''(r)$ but not higher derivative terms such as $A''(r)^2$, $A'(r) A^{(3)}(r)$, or $A^{(4)}(r)$ appear in $T^{t}_{\ t} - T^{r}_{\ r}$ after using the higher derivative modified equations of motion. In our higher derivative gravity \eqref{higher} with the metric ansatz \eqref{holoren}, we have
\begin{align}
I^{t}_{\ t} - I^{r}_{\ r} &= 72 (A''(r))^2 + 36 A'(r) A^{(3)}(r) + 12 A^{(4)}(r)  \cr
J^{t}_{\ t} - J^{r}_{\ r} &=  24 (A''(r))^2 + 12A'(r) A^{(3)}(r) + 4 A^{(4)}(r) \cr
\end{align}
so that the requirement is equivalent to $\lambda_2 + 3\lambda_3 = 0$, which is nothing but $W_{\mu\nu\rho\sigma}^2$ term up to a total derivative. Note that as we have already mentioned, we a priori knew that this combination does not modify the equations of motion for conformally Einstein manifold from the one in Einstein gravity. 

Under this setup, we can easily show that the holographic $c$-function
\begin{align}
a(r) \equiv  \frac{\pi^{3/2}}{\Gamma(3/2)(A'(r))^2} + a_0 \ 
\end{align}
is monotonically decreasing along the holographic renormalization group flow as long as the null energy condition is satisfied in this higher derivative gravity. The constant $a_0$ can depend on the topological terms and the relation to the entropy-like formula suggests its necessity (see \cite{Myers:2010tj} for more details) although it is irrelevant for the holographic renormalization group flow. It is curious, however, when the theory confines, $A'(r)$ becomes effectively infinite, but with non-zero $a_0$, we still have non-zero $c$-function (and it can be negative).

A further analysis of the linearized perturbation shows that the absence of the higher derivative terms than $A''(r)$ in  $T^{t}_{\ t} - T^{r}_{\ r}$ is necessary to guarantee the absence of a ghost mode in higher derivative gravity along the holographic renormalization group flow. Actually, the sufficient condition seems to require a slightly stronger condition that $\lambda_2 = \lambda_3 = 0$ \cite{Myers:2010tj}. We have not seen this because $\lambda_2 + 3\lambda_3 = 0$ condition only guarantees the absence of a ghost within the deformation that is conformally Einstein. If we demanded the stronger condition, there would be no allowed non-topological curvature squared corrections in $(1+3)$ dimension, but for the validity of the holographic $c$-theorem, we may not have to impose it (see  also \cite{Liu:2010xc}\cite{Liu:2011iia} for a more relaxed condition on the higher derivative holographic $c$-theorem).

In the above argument, it was crucial that the matter sector satisfies the null energy condition. At this point, we should have worried about it because due to the quantum effect, the null energy condition can be easily violated. In particular, as we will discuss in the next section, a contribution from 1-loop anomalous terms in energy-momentum tensor is typically of order $\mathcal{O}(R^2)$. Therefore, there is no point in discussing the higher curvature corrections in $(1+3)$ dimensional bulk gravitational theory without considering the quantum corrections to the energy-momentum tensor and its possible violation of the null energy condition. Conversely, the counterterms needed for the renormalization of the quantum corrected energy-momentum tensor are nothing but the curvature terms added in \eqref{higher}, so we cannot talk about the fate of the holographic renormalization group flow by not considering both at the same time. The only excuse would be taking the large $N$ limit of AdS/CFT correspondence, in which $\alpha'$ corrections and $g_s$ corrections can be separated. We will rather work on the effective field theory approach where there is no obvious distinction of the two with more generic holographic duality in mind, but we will have a small comment on the string $g_s$ correction in section 4.

\section{3. Anomalous Violation of Null Energy Condition and Fate of $c$-theorem}
The null energy condition can be violated in various ways in quantum mechanics. We will focus on one specific but universal violation induced by the trace anomaly. Most of the other reported violations break the time-translation, so they are not of our immediate interest. The anomalous violation of the null energy condition here is tightly related to the trace anomaly. Suppose we have a conformal field theory in $(1+3)$ dimension. By its definition, in flat Minkowski space-time, the energy-momentum tensor is traceless
\begin{align}
T^{\mu}_{\ \mu} = 0 \ ,
\end{align}
but in generic space-time, it shows a trace anomaly (see e.g. \cite{Deser:1993yx}\cite{Duff:1993wm}\cite{Birrell:1982ix} and references therein)
\begin{align}
{T}^{\mu}_{\ \mu} = a \mathrm{Euler} -c W_{\mu\nu\rho\sigma}^2 + b' \Box R + e \epsilon^{\alpha\beta \gamma \delta} R_{\alpha \beta \rho \sigma} R_{\gamma \delta}^{\ \ \rho \sigma} 
\end{align}
Here for completeness, we have introduced a possible CP-odd term $\epsilon^{\alpha\beta \gamma \delta} R_{\alpha \beta \rho \sigma} R_{\gamma \delta}^{\ \ \rho \sigma}$ \cite{Nakayama:2012gu}, but the effect (if any) will vanish on conformally flat space-time we are interested in. 

We will not need the actual number such as $a$ and $c$. We only note that $c$ is manifestly positive for unitary conformal field theories due to the positivity of the two-point functions of the energy-momentum tensor. The positivity of $a$ is more delicate, but we believe that it is the case (strong evidence can be found in \cite{Hofman:2008ar} from the averaged null energy condition in flat Minkowski space-time).
 In contrast, the number $b'$ is ambiguous because we can always change its value by adding the local counterterm $R^2$ to the action. It is a common tradition that we do not call it anomaly when it can be removed by a local counterterm.

Given the trace-anomaly, the energy-momentum tensor is no longer a conformal primary tensor, and for conformal field theories, the conformal transformation induces the correction terms 
\begin{align}
\bar{T}^{\mu}_{\ \nu} &= \Omega^{-4} T^{\mu}_{\ \nu}  \cr
 &- 8c \Omega^{-4}\left((W^{\alpha \mu}_{\ \ \beta \nu}\log \Omega)^{;\beta}_{;\alpha} + \frac{1}{2}R^{\beta}_{\alpha} W^{\alpha\mu}_{\ \ \beta \nu} \log \Omega \right)\cr
&-a \left(4\bar{R}^{\beta}_{\alpha} \bar{W}^{\alpha \mu}_{\ \ \beta \nu} -2 \bar{H}^{\mu}_{\nu} - \Omega^{-4}(4R^{\beta}_{\alpha} W^{\alpha \mu}_{\ \ \beta \nu} - 2H^{\mu}_{\nu}) \right) \cr
& -\frac{1}{6}b' \left(\bar{I}^{\mu}_{\nu} - \Omega^{-4} I^{\mu}_{\nu} \right) \ , \label{anoma}
\end{align}
where
\begin{align}
H_{\mu\nu} = -R^{\alpha}_{\mu} R_{\alpha \nu} + \frac{2}{3} R R_{\mu\nu} + \left(\frac{1}{2} R_{\alpha \beta} R^{\alpha \beta} - \frac{1}{4}R^2 \right) g_{\mu\nu} 
\end{align}
as first proposed in \cite{Page:1982fm}. Here $\bar{g}_{\mu\nu} = \Omega^2 g_{\mu\nu}$, and barred quantities are evaluated by using $\bar{g}_{\mu\nu}$. 
Note that the right hand side is covariantly conserved  and the trace reproduces the trace anomaly, so it is a consistent covariant quantity that can be used in diffeomorphism invariant equations of motion.

For the reason we will explain later, we will set $b'=0$ by adding a suitable counterterm as mentioned above. Then in the conformal flat space-time we are interested in, the anomalous contribution to conformal transformation of the energy-momentum tensor is simplified:
\begin{align}
\bar{T}_{\mu\nu} = \Omega^{-2} T_{\mu\nu} -2a \left(\bar{R}_{\mu}^{\rho} \bar{R}_{\rho \nu} - \frac{1}{3}\bar{R} \bar{R}_{\mu\nu}\right) \ .
\end{align}
Our expression is different from that in \cite{Urban:2009yt}\cite{Urban:2010vr}  because we have set $b'=0$ by introducing the counterterm while they used $b'=\frac{2}{3}c$ motivated by the dimensional regularization.

As a digression, we now understand we can easily violate the null energy condition with this quantum correction even in a conformal vacuum \cite{Urban:2009yt}\cite{Urban:2010vr} . Suppose we have a conformally flat metric with small conformal transformation $\omega = \log \Omega \ll 1$ in the initial flat Minkowski vacuum state $T_{\mu\nu} = 0$. With the light-cone coordinate $z = x-t$ and $\bar{z} = x+t$, we have the anomalous contribution to the conformally transformed energy-momentum tensor
\begin{align}
\bar{T}_{zz} = -8a \Omega^{-2} g^{\mu\nu} \partial_z \partial_\mu \omega \partial_z \partial_\nu \omega + \mathcal{O}(\omega^3) \ .
\end{align}
There is no difficulty in finding $\omega$ that gives a negative value for $\bar{T}_{zz}$ by choosing a space-like vector $\partial_z \partial_\mu \omega$ (see e.g. \cite{Urban:2010vr}  for a choice). The violation persists even after averaging over a null  geodesics.

Coming back to our problem, we start with the AdS invariant vacuum, where $T_{\mu\nu} \propto g_{\mu\nu}$, and do the conformal transformation to generate the anomalous contribution to the energy-momentum tensor in the holographic renormalization group flow metric \eqref{holoren}. By itself, the null energy condition is violated in the anomalous contribution 
\begin{align}
(\bar{T}^{r}_{ \ r}-\bar{T}^t_{\ t} )|_{\mathrm{anom}} &= 2a(\bar{H}^r_{\ r}-\bar{H}^t_{\ t}) \cr
&= 4aA''(r)(A'(r))^2 \le 0 \ \label{vio}
\end{align}
because as we will see $A''(r) \le 0$ in the consistent background despite the violation here. We emphasize that the self-consistency condition is not necessarily imposed when we say that the null energy condition is violated. 

Nevertheless, the holographic $c$-theorem is actually intact in the full self-consistent holographic renormalization group flow with a suitable modification of the holographic $c$-function. We will consider the gravitational action that may contain the curvature squared corrections but with no higher derivative corrections in linearized equations of motion as discussed in the last section (i.e. beside the topological term, we only allow $W_{\mu\nu\rho\sigma}^2$ term).
The equations of motion for the warp factor now demands
\begin{align}
2A''(r) = ({T}^{t}_{\ t} - {T}^{r}_{\ r}) - 4a A''(r)(A'(r))^2 \ ,
\end{align}
where $T_{\mu\nu}$ here is the non-anomalous (classical) part of the matter energy-momentum tensor in the holographic renormalization group flow geometry.
We can define the suitably modified holographic $c$-function as
\begin{align}
a(r) \equiv  \frac{\pi^{3/2}}{\Gamma(3/2)(A'(r))^2} - 4a\frac{\pi^{3/2}}{\Gamma(3/2)} \log A'(r) \ ,
\end{align}
which is indeed monotonically decreasing
\begin{align}
a'(r) =  - \frac{\pi^{3/2}}{\Gamma(3/2) (A'(r))^3} (T^{t}_{\ t} - T^{r}_{\ r}) \ge 0 \ 
\end{align}
as long as the classical part of the energy-momentum tensor $T_{\mu\nu}$ satisfies the classical null energy condition: $T^{t}_{ \ t} - T^{r}_{\ r} \le 0$ so that $A''(r)  \le 0$ (recall that $a$ is positive in unitary conformal field theories).

In this way, we have shown that the possible violation of the null energy condition by the anomalous terms in energy-momentum tensor, which we believe universal, is not an immediate threat to the holographic $c$-theorem as long as we eliminate $b' \Box R$ term in the trace anomaly by adding a suitable counterterm. Actually, this term would have spoiled the above argument because as discussed in section 3, it would have introduced higher derivative terms such as $A''(r)^2$, $A'(r)A^{(3)}(r)$ or $A^{(4)}(r)$ in $\bar{T}^{t}_{ \ t} - \bar{T}^{r}_{ \ r}$ through the anomalous contribution to the energy-momentum tensor (i.e. from the last term  with $\bar{I}_{\mu\nu}$ in \eqref{anoma}) and the monotonicity of the holographic $c$-function becomes non-obvious \cite{Myers:2010tj}\cite{Liu:2010xc}\cite{Liu:2011iia}. In retrospect, this term ends up with introducing a ghost mode in the consistent equations of motion as is obvious from the higher derivative terms appearing in the evolution of the warp factor, which had better be avoided. This essentially teaches us that in order to discuss the consistency of the background with $\mathcal{O}(R^2)$ corrections, we have to study the curvature squared corrections to the action as well as the quantum corrections to the energy-momentum tensor simultaneously. 

In contrast, the $\log$ correction we found in the holographic $c$-function cannot be removed by adding any local counterterm of order $\mathcal{O}(R^2)$ or higher to the effective action. Therefore, its contribution is universal. This is expected because the origin of the $\log$ correction is the trace anomaly and one cannot remove it by any local counterterm. It is assuring that the holographic $c$-theorem is still intact nevertheless.

\section{4. Conclusion}
In this paper, we have studied how a universal violation of the null energy condition from trace anomaly affects the holographic renormalization group flow in $(1+3)$ dimension. We stress again that the discussion of the quantum corrections in energy-momentum tensor is absolutely necessary at the curvature squared level before we talk about any higher curvature corrections such as $R^3$ added in the action. The contribution we discussed is particularly of interest because it is non-local and cannot be attributed to a variation of any local effective action.

To avoid the confusion, we should clarify the terminology of the violation of the null energy condition here. What we mean by the violation of the null energy condition due to the anomalous part of the energy-momentum tensor is conventional. We fix the background (e.g. \eqref{holoren}) by hand, and compute the anomalous matter energy-momentum tensor as if there were no other fields. Then as discussed around \eqref{vio}, the null energy condition is violated. These are the reported null energy violations in the literature.

However, as we saw in what was happening in a suitably modified holographic $c$-theorem argument, this violation is not so relevant. What is important to us is the self-consistent background that solves the total gravitational equations of motion derived from the anomalous energy-momentum tensor. With this respect, we can say that the ``self-consistent" null energy condition is not violated in our example even with the anomalous contribution. The reason is that we needed a classical background that produces the holographic renormalization group flow metric \eqref{holoren}, and the classical part is far larger than the quantum negative contribution (as far as $a$ is positive).

Nevertheless, we have learned some lessons. First of all, we had to add a suitable counterterm to cancel a possible ghost mode that could appear in the self-consistent background. We note that would-be natural value for $b'$ is $\frac{2}{3}c$, which is given in dimensional regularization. However, we showed that in order to obtain the holographic $c$-theorem, this regularization is problematic. Of course,  we can simply conclude that the dimensional regularization here is not consistent with unitarity, but we would like to understand a more detailed physical explanation why this is the case.

It would be of great interest if we could answer this regularization issue in direct string theory one-loop computation. Unfortunately, the only reliable computation are made in perturbation around the Minkowski space-time, and although we have discovered the on-shell trace anomaly and associated anomalous energy-momentum tensor in flat-Minkowski space \cite{Antoniadis:1992sa}, this computation does not tell whether $b' \Box R$ term is absent or not because it vanishes on-shell around the Minkowski space-time.

Another interesting observation is that the appearance of $\log A'(r)$ in the holographic $c$-function. This means that unlike the holographic $c$-function for boundary field theories in even dimensions, there is no absolute meaning in the actual value in the $c$-function for those in odd dimensions with the anomalous energy-momentum tensor in the bulk gravity. As a corollary, there is no absolute meaning in the positivity of the odd-dimensional holographic $c$-function here. It must be contrasted with the positivity of the central charge $c$ in $(1+1)$ dimensional unitary conformal field theories, as well as the positivity of $c$ and $a$ in $(1+3)$ dimensional unitary conformal field theories.

Without the anomalous contribution from the energy-momentum tensor, the value of the holographic $c$-function is identified with the universal contributions to the entanglement entropy or the sphere free energy in three-dimension \cite{Myers:2010tj}\cite{Casini:2011kv}. Our modification suggests that the holographic computation of the entanglement entropy must be modified accordingly. This seems to be a difficult question to ask, but the recent computation of the one-loop supergravity corrections to the free-energy from holography \cite{Bhattacharyya:2012ye} is encouraging, in which $\log N$ term was found as one-loop corrections. Since the corresponding field theories do possess the similar $\log$ corrections in the free energy in large $N$ limit, we believe that our prescription of the holographic $c$-function is consistent with the field theory conjecture.
 With this respect, it is curious to reconcile with the conjecture that the three-dimensional $c$-function, which is  proposed as the finite part of the sphere free energy, is positive \cite{Jafferis:2011zi}.

Finally, at various technical point, it was crucial that our background preserves the Poincar\'e invariance in $d$-dimension. This makes the holographic renormalization group background conformally flat, and the anomalous contribution to the energy-momentum tensor is simplified a lot. In particular, we note that only $a$ anomaly appears but $c$ does not. This is no longer the case once we abandon the Poincar\'e invariance. There is very few understanding of non-relativistic version of the holographic $c$-theorem, let alone field theory understanding. Probably, the violation of the null energy condition can be more drastic there.

The null energy condition appears in other areas of holography such as boundary $g$-theorem \cite{Fujita:2011fp}\cite{Nakayama:2012ed}, the limit on the dynamical critical exponent \cite{Hoyos:2010at}\cite{Ogawa:2011bz}\cite{Dong:2012se}\cite{Liu:2012wf} and so on.
We should ask similar questions there. Perhaps, rather than studying each problem, it would be better to understand the whole picture by asking ``what will effectively replace the null energy condition in quantum gravity".

\section*{Acknowledgments}
The work is supported by Sherman Fairchild Senior Research Fellowship at California Institute of Technology.


\end{document}